\documentclass{conf_en}

\usepackage[T2A]{fontenc}
\usepackage[cp1251]{inputenc}
\usepackage[russian]{babel}
\usepackage{array}

\title{{\bf CANONICAL FORMALISM OF BIGRAVITY}\footnote{``Black and Dark Topics of Modern Cosmology and Astrophysics'', Dubna, September 15-22, 2013.}}
\author{Vladimir O. Soloviev\footnote{IHEP, Protvino, 142281, Ploschad' Nauki, 1.}\footnote{International University ``Dubna'', Filial "Protvino", 142281, Severnyj Proezd, 9.}}

\begin{document}
  \maketitle
  \begin{abstract}
 We construct the Hamiltonian formalism of bigravity for the  potential of a general form. We find 
conditions on this potential and prove that under these conditions the formalism is equivalent to the one constructed with the celebrated dRGT-potential. 
  \end{abstract}
\section{Introduction}
One way to explain the observable accelerated expansion of the Universe is to modify the theory of gravitation on the cosmological scale. Here the massive gravity~\cite{Reviews} and the bigravity are among the most popular generalizations of the General Relativity (GR).

For a long time it was believed that one could not avoid the Boulware-Deser (BD) ghost~\cite{BD} in massive gravity. But recently a new model free of this ghost has appeared~\cite{dRGT}. The skeptics are still sure that any massive gravity is inconsistent, and their new analysis discovers superluminal effects and causality violations~\cite{Deser}. But it is not excluded for sure that these difficulties could be avoided in a theory with two (or even more) dynamical metrics, i.e. in the bigravity (or multigravity). Such a theory first appeared at the beginning of 1970s~\cite{ISS}, the interest to it has been periodically dissipating and  arising. The next wave has come at the Millenium~\cite{Kogan}, related with the extra-dimensions and with the evidence for the accelerated expansion of the Universe. And the recent activity is inspired by the discovery of dRGT-potential with its interesting properties.
(Curiously, it was recently mentioned that one case of this potential had already been under study 40 years ago~\cite{Zumino}.)

\section{Hamiltonian formalism}
The analysis of theory structure can be provided in the canonical formalism without any perturbation theory. The first research was parallel to the Hamiltonian study of the massive gravity and  based on the transform of variables found by Hassan and Rosen~\cite{HR}. Straightforward calculations are unavailable due to the unusual form of dRGT-potential. It is a linear combination of the symmetric polynomials of matrix $\mathbf{Y}$ eigenvalues. This matrix in its turn is a square root of matrix $\mathbf{Z}^\alpha_\beta=g^{\alpha\mu}f_{\mu\beta}$. It is unclear how to find derivatives of such a potential,  so, it is difficult to derive the constraints and the Hamiltonian formalism as a whole.

This problem has been solved in Ref.~\cite{HR} with the help of a special transform of variables and with nontrivial matrix calculation. We believe there are no more than 5-10 physicists in the world who could reproduce this proof, and it is useless to try teaching it to students. Therefore we have tried to construct a more transparent approach which allows to do the standard procedures with the appearing constraints: to calculate the Poisson and Dirac brackets, to separate the constraints into first and second class sets, to prove the closure and self-consistence of the total set of constraints and the Hamiltonian.

The idea was to take a theory with the potential of a general form depending on the two space-time metrics, or to be more precise, on the components of their $3+1$-decomposition. Whereas in the Arnowitt, Deser, Misner (ADM) formalism~\cite{ADM} the metrics are decomposed in the coordinate basis 
 $$
||f_{\mu\beta}||=
\left(
\begin{array}{cc}
-N^{2}+N^kN_k & N_j\\
N_i & \eta_{ij}
\end{array}\right), \ ||g^{\alpha\mu}||=
\left(
\begin{array}{cc}
-{\bar N}^{-2} & {\bar N}^j{\bar N}^{-2}\\
{\bar N}^i{\bar N}^{-2} & \gamma^{ij}-{\bar N}^i{\bar N}^j{\bar N}^{-2}
\end{array}\right),
$$
we prefer the semi-orthonormal basis constructed on the base of metric $f_{\mu\nu}$. This formalism first developed by Kucha\u{r}~\cite{Kuchar} is more suitable for the bigravity because a number of components for metric $f_{\mu\nu}$ is reduced from ten to six:
$$
||f_{\mu\beta}||=
\left(
\begin{array}{cc}
-[n_\mu n_\beta] & 0\\
0 & \eta_{ij}[e_\mu^i e_\beta^j]
\end{array}\right), \ ||g^{\alpha\mu}||=
\left(
\begin{array}{cc}
-u^{-2}[n^\alpha n^\mu ] & u^ju^{-2}[n^\alpha e^\mu_j]\\
u^iu^{-2}[e^\alpha_i n^\mu] & (\gamma^{ij}-u^iu^ju^{-2})[e^\alpha_i e^\mu_j]
\end{array}\right),
$$
this considerably simplifies the potential which now depends on sixteen variables, instead of twenty.

The Kucha\u{r} method exploits two coordinate systems for the space-time: one, $X^\alpha$, is arbitrary, the other $(\tau,x^i)$ is related to the choice of a one-parametric family of space-like  hypersurfaces. The second system coincides with the one used in the ADM approach.  The four functions called embedding variables $X^\alpha=e^\alpha(\tau,x^i)$ give a one-to-one map. In fact these variables appeared already in the pioneer works by Dirac~\cite{Dirac} where the diffeomorphism invariant field theory Hamiltonian approach was proposed.

The diffeomorphism invariance of the canonical formalism manifests itself as a path independence of  evolution. It means that in evolving from an initial spatial hypersurface to a final one the result should be independent on the intermediate sequence of hypersurfaces. As it has been shown by Teitelboim~\cite{Teit} in order to fulfill this requirement it is necessary and sufficient to have a special algebra of the first class constraints,  first calculated by Dirac.  Starting from the potential of a general form, we  derive the {\bf first conditions} on the potential in order to get this algebra in bigravity. These are linear differential equations of the first order.  

Next, the simple calculation shows that if we have only four first class constraints then we come to a theory of bigravity with eight degrees of freedom, so containing BD ghost~\cite{BD}. Therefore the second requirement on the potential comes from the need to provide an additional constraint giving us a chance to avoid this ghost. This should be a constraint on the fundamental canonical variables of bigravity, i.e. two spatial metrics (induced on hypersurfaces) $\eta_{ij}=f_{\mu\nu}e^\mu_ie^\nu_j$, $\gamma_{ij}=g_{\mu\nu}e^\mu_ie^\nu_j$ and their conjugate momenta $\Pi^{ij}$, $\pi^{ij}$,  twelve pairs of conjugate variables in total. The theory also contains eight non-dynamical variables $N,N^i$, $u,u^i$, and the Lagrangian does not contain their time derivatives. If we vary the Lagrangian with respect to $N,N^i$, we get the above mentioned four first class constraints
$$
{\cal R}=0,\qquad {\cal R}_i=0,
$$
and if we vary it with respect to $u,u^i$, we get four other equations
\begin{equation}
{\cal S}=0,\qquad {\cal S}_i=0,\label{eq:SSi}
\end{equation}
which serve to determine  $u,u^i$ in terms of the canonical variables. To change this situation one needs a degeneracy not allowing to resolve Eqs.(\ref{eq:SSi}) for all  $u,u^i$. Therefore the Jacobian 
 $\frac{D ({\cal S},{\cal S}_i)}{D (u,u^j)}$ should be zero.
Due to a special form of Eqs.(\ref{eq:SSi}) this Jacobian occurs to be equal to the Hessian of the potential considered as a function of $u,u^i$. Then the {\bf second requirement} put on the potential is to fulfill the Monge-Amp\`ere equation
$$
\left|\frac{\partial^2 \tilde U}{\partial u^a\partial u^b}\right|=0.
$$
Then what rank should the Hessian matrix have? We believe it should be equal to three. This is the {\bf third condition} on the potential. Taking a smaller number will exclude isotropic spaces from the very beginning.  

We found that the three requirements listed above (of course, there is also a {\bf zeroth condition} -- that the potential is a function of metric components) are sufficient to have a diffeomorphism invariant theory of bigravity which is free of BD ghost, i.e. it has seven gravitational degrees of freedom. It is natural to expect two species of matter in bigravity: each matter interacts with the corresponding metric. There is a picture of two extremely weakly interacting worlds where only the two metrics directly interact with each other.

The proof of sufficiency of the proposed conditions on the potential consists in the analysis of constraints by the well-known Dirac's method. It is given in detail in articles~\cite{SolTch}. In the course of this work the implicit solution of the Monge-Amp\`ere equation constructed by Fairlie and Leznov~\cite{Leznov} is very useful, as this has been first demonstrated in the study of massive gravity with one dynamical metric~\cite{Comelli}.

\section{Conclusion}
The following results are proved for the theory of bigravity with the potential satisfying conditions given above.
\begin{itemize}
\item Besides ${\cal R}$ and ${\cal R}_i$ there is an additional fifth constraint on canonical variables ${\cal S}=0$, it is free of non-dynamical variables $u,u^i$.

\item  ${\cal S}$ has a zero Dirac bracket with itself (on the constraint surface). 

\item To preserve  ${\cal S}=0$ in the course of evolution it is necessary to have sixth constraint $\Omega=0$, also free of non-dynamical variables.

\item The Dirac bracket 
$\{{\cal S},\Omega\}_D$ is nonzero, so these two constraints are second class.

\item To preserve  $\Omega=0$ in the course of evolution it is necessary to fulfill a new equation $\Psi=0$, which is a linear inhomogeneous equation on the non-dynamical variable  $u$.  This variable can be  determined from $\Psi=0$.~\footnote{Two cases are possible here:   $u$ is expressible in canonical variables only, or it also depends on Lagrangian multipliers $N,N^i$. 
Anyhow, these Lagrangian multipliers for the first class constraints are arbitrary.  This persuades us in the diffeomorphism invariance of bigravity.}

\end{itemize}
{\small


\begin{thebibliography}{**}

\bibitem{Reviews}
V.A.~Rubakov and P.G.~Tinyakov,  {\it Phys.-Uspekhi}\  {\bf 51} 759-822 (2008); arXiv:0802.4379;
 D. Blas,
 Aspects of Infrared Modifications of Gravity;
arXiv:0809.3744;
K.~Hinterbichler,
{\it Rev. Mod. Phys.} {\bf 84} 671-710 (2012); arXiv:1105.3735.

\bibitem{BD}
D.G. Boulware and S. Deser, {\it Phys.Rev. } {\bf D 6} 3368-3382 (1972). 

\bibitem{dRGT}
C. de Rham, G. Gabadadze, and A. J. Tolley,
{\it Phys. Rev. Lett.} {\bf 106} 231101 (2011); 	arXiv:1011.1232;
{\it Phys. Lett. B}
{\bf 711} 190-195 (2012); arXiv:1107.3820

\bibitem{Deser}

S.~Deser and A.~Waldron, 
{\it Phys. Rev. Lett.} {\bf 110} 111101 (2013); arXiv:1212.5835.

\bibitem{ISS} 
C.~J.~Isham, A.~Salam, and J.~Strathdee, {\it Phys. Lett.} {\bf B 31} 300-302 (1970); {\it Phys. Rev.} {\bf D 3} 867-873 (1971).

\bibitem{Kogan}
T. Damour and I.I. Kogan,
{\it Phys.Rev. } {\bf D 66}  104024 (2002).







\bibitem{Zumino} B.~Zumino, ``Effective Lagrangians and broken symmetries,''
  in Brandeis Univ. Lectures on Elementary Particles and Quantum Field Theory (MIT Press Cambridge, Mass.), Vol. 2,  1970, p.437.


\bibitem{HR}
S. F. Hassan and R.A. Rosen, 
{\it Phys. Rev. Lett.} {\bf 108} 041101 (2012);	arXiv:1106.3344;
	{\it JHEP}
{\bf 1202} 126 (2012);	 	 	arXiv:1109.3515;
{\it JHEP} {\bf 1204}  123 (2012);
	arXiv:1111.2070;
S. F. Hassan, R.A. Rosen, and A. Schmidt-May,
	 	{\it JHEP} {\bf 1202} 026 (2012); arXiv:1109.3230.

\bibitem{ADM}
R.~Arnowitt, S.~Deser, and Ch.W.~Misner, in  Gravitation, an Introduction
to Current Research, ed. L.~Witten,  Wiley, New York (1963); arXiv:gr-qc/0405109. 

\bibitem{Kuchar}
 K. Kucha\u{r},  {\it J.~Math.~Phys.} 
 {\bf 17} 777-791; 792-800; 801-820 (1976);
{\bf 18} 1589-1597 (1977).

\bibitem{Dirac}
P.A.M. Dirac,  Lectures on Quantum Mechanics. Yeshiva University, New York, (1964). 

\bibitem{Teit}
C. Teitelboim, 
{\it Ann. Phys. (N.Y.)} {\bf 79} 542-557 (1973).



\bibitem{SolTch}
V.O. Soloviev and M.V. Chichikina, {\it Teoret. Mat. Fiz.}
{\bf 176}, 393-407 (2013) [{\it Theoret. Math. Phys.} {\bf 176}, 1163-1175 (2013)];
arXiv:1211.6530; V.O. Soloviev and M.V. Tchichikina,
{\it Phys. Rev.} {\bf D 88} 084026 (2013);
arXiv:1302.5096.

\bibitem{Leznov}
D. Fairlie and A. Leznov.
{\it J. Geom. Phys.} {\bf 16} 385-390 (1995); arxiv:hep-th/9403134.

\bibitem{Comelli}
D. Comelli, F. Nesti, and L. Pilo, {\it Phys. Rev.} {\bf D 87} 124021 (2013);
 arXiv:1302.4447;
{\it JHEP} {\bf 1307} 161 (2013);
arXiv:1305.0236.












 












\end{thebibliography}
\end{document}